\documentclass[a4paper,10pt]{article}

\pdfoutput=1

\usepackage[toc,page]{appendix}

\usepackage{rotating}
\usepackage{listings}

\usepackage{amsmath}
\usepackage{amssymb}
\usepackage{amsthm}
\usepackage{mathpartir}

\usepackage[utf8]{inputenc}

\usepackage[T1]{fontenc}

\usepackage[backend=bibtex]{biblatex}

\usepackage{url}
\bibliography{bibliography.bib}

\newlength\digitwidth
\footnotesize
\let\LSTINLINE\lstinline
\def\lstinline{\LSTINLINE[basicstyle=\sffamily\upshape]}

\renewcommand*\thelstnumber{\ifnum\value{lstnumber}<10\relax\kern\digitwidth\fi\arabic{lstnumber}}

\makeatletter
\renewenvironment{figure}
  {\@float{figure}\footnotesize}
  {\end@float}
\makeatother

\let\phi\varphi

\let\variables V

\newcommand*{\mkterm}[1]{\mathsf{#1}}

\newcommand{\reduces}{\longrightarrow}

\newcommand*{\subs}[2]{\{^{#1}\!{/}_{#2}\}}
\newcommand*{\stat}[2]{({#1},{#2})}

\newcommand{\bnf}{\;::=\;}
\newcommand{\alt}{\;\;|\;\;}
\newcommand\classterm{\mkterm{class}}

\newcommand*{\class}[4]{\mkterm{class}~{#1}~\{#2;#3;#4\}}

\newcommand*{\methodd}[4]{{#1}~{#2}(#3)~\{#4\}}

\usepackage{color}

\newcommand\this{\mkterm{this}}

\newcommand\methsterm{\mkterm{methods}}
\newcommand\fieldsterm{\mkterm{fields}}

\newcommand\usageterm{\mkterm{usage}}

\newcommand*{\methcal}[3]{{#1}.{#2}({#3})}

\newcommand*{\selfcal}[2]{{#1}({#2})}

\newcommand*{\seq}[2]{{#1};{#2}}
\newcommand*{\der}[2]{{#1}.{#2}}
\newcommand*{\upd}[3]{{\der{#1}{#2}}={#3}}

\newcommand{\whileterm}{\mkterm{while}}

\newcommand*{\while}[2]{\whileterm~({#1})\{#2\}}

\newcommand{\ifterm}{\mkterm{if}}
\newcommand{\elseterm}{\mkterm{else}}
\newcommand*{\cond}[3]{\ifterm~({#1})~#2~\elseterm~#3}

\newcommand{\ectx}{\mathcal{E}}

\newcommand\nullterm{\mkterm{null}}

\newcommand\unit{\mkterm{unit}}
\newcommand\unittype{\mkterm{void}}

\newcommand\booltype{\mkterm{bool}}
\newcommand\true{\mkterm{true}}
\newcommand\false{\mkterm{false}}
\newcommand*{\rectype}[2]{\mu{#1}.{#2}}
\newcommand*{\objecttype}[2]{{#1}[#2]}

\newcommand*{\echoicetypel}[4]{\{#1_{#3};#2_{#3}\}_{#3\in #4}}

\newcommand\linterm{\mkterm{lin}}
\newcommand\unterm{\mkterm{un}}

\newcommand*{\typedexp}[4]{{#1}\vartriangleright{#2} : {#3}\vartriangleleft{#4}}
\newcommand*{\typedexpc}[4]{{#1}\vartriangleright_{C}{#2} : {#3}\vartriangleleft{#4}}
\def\judgment#1>#2:#3<#4/{\ensuremath{\typedexp{#1}{#2}{#3}{#4}}}

\def\subjudgment#1>#2:#3<#4/{\ensuremath{{#1}\triangleright{#2}\subt{#3}\triangleleft{#4}}}
\def\judgmentc#1>#2:#3<#4/{\ensuremath{\typedexpc{#1}{#2}{#3}{#4}}}

\newcommand{\typedclass}[1]{\vdash #1}

\newcommand{\typedval}[3]{#1 \vartriangleright #2: #3}

\newcommand*\changetype[3]{{#1}\{#2\mapsto #3\}}
\newcommand*\changeval[3]{{#1}\{#2\mapsto #3\}}

\newcommand{\subt}{\mathrel{\mathsf{<:}}}

\newcommand{\allowsterm}{\mkterm{allows}}

\newcommand{\agreeterm}{\mkterm{agree}}

\newcommand*{\comment}[1]{}
\newcommand\ignore[1]{}
\makeatletter\let\noqed\@qededtrue\makeatother

\newcommand{\defeq}{\stackrel{\mathrm{def}}{=}}

\lstdefinelanguage{spi}
 {morekeywords=[1]{ok,pair,fst,snd,Un,Key},
  morekeywords=[1]{Ok,Pair,Ch,empty,union},
  morekeywords=[2]{out,in,new,event,if,then,else,let,else,begin,end,exercise,decrypt,as,split,match},
  morekeywords=[3]{TermCtor,TermData,TermDtor,Predicate,Process,Private,TopLevel,Main},
  sensitive=true,%
  morecomment=[l]{//},
  literate={->}{$\to\:$}3,
  literate={<=}{$\leq\:$}3,
  morestring=[b]"}

\newcommand{\Tenv}{\Gamma}

\title{A Revision of the Mool Language}
\author{Cláudio Vasconcelos \qquad António Ravara\\\\
  NOVA-LINCS and Dep. de Informática, FCT.\\
  Universidade NOVA de Lisboa, Portugal}

\begin{document}
\maketitle

\begin{abstract}
  We present here in a thorough analysis of the
  Mool language, covering not only its implementation but also the
  formalisation (syntax, operational semantics, and type system).
  The objective is to detect glitches in both the implementation and
  in the formal definitions, proposing as well new features and added
  expressiveness.
  To test our proposals we implemented the revision developed in the
  Racket platform.
  % This article presents a analysis to the public formalization of the
  % Mool language. The objective is to detect where the language is not
  % well defined and propose a revision of the language. To test our
  % revision we implemented our revision of the Mool language
  % formalization in Racket.
\end{abstract}

\section{Introduction}
This article presents an analysis of the Mool language, a small
object-oriented language similar to Java, developed by Campos and
Vasconcelos~\cite{DBLP:journals/corr/abs-1110-4157,joanacorreiacampos2010}.
The language allows to associate with each class a behavioural type
specifying safe orderings of method calls, along the lines of~\cite{modular}.

This analysis is a contribution to the development of the language,
detecting bugs not only in the implementation, but also in
formalisation. We also propose revisions of aspects of the language we
find too restrictive.
% This article consists on the analysis of the Mool language, a small
% object-oriented language similar to Java, presented by Campos in
% \cite{joanacorreiacampos2010}, that allows to associate with each
% class a behavioral type specifying safe orderings of method
% calls. With this analysis we pretend to detect were the language, as
% presented in its public formalization, is not well defined, i.e., it
% contains an error that has influence on the language behaviour or it
% is too restricting.

Section \ref{sec:mool_problems} presents correction proposals. We
organise them in two categories: minor aspects (Section
\ref{sec:mool_problems_minor}), which have little influence on the
language or their correction is very straightforward; and major
aspects in(Section \ref{sec:mool_problems_major}), which heavily
influence the behaviour of the language and are more complex to
change.
% Section \ref{sec:mool_problems} presents the aspects where we think the language should be corrected. We categorised it in two separate categories, with the aspects in section \ref{sec:mool_problems_minor} being the minor aspects that have little influence on the language or its solution is very straightforward, and the aspects in \ref{sec:mool_problems_major} being the most important ones because they heavily influence the language and are more complex to solve.

We complement the analysis of the Mool language formal system with a
small review of the Mool compiler (version 0.3, available in May
2016 from \url{gloss.di.fc.ul.pt/mool/download}). The purpose is
to understand if the aspects we presented in section
\ref{sec:mool_problems} were solved in the implementation, and, if
they were, how the compiler copes with them.
% We complement this analysis on the original formalization of the Mool language with a small review of the latest version of the Mool language available. The purpose of this review was to try to understand if the aspects we presented in section \ref{sec:mool_problems} were solved and, if they were, how the compiler copes with them.

To test our analysis, we implemented the original formalisation of
Mool using PLT-Redex \cite{redex}, a module available in Racket
\cite{manifesto} that allows us to implement and debug formal systems
of programming languages. Section \ref{sec:mool_redex_1} present our
implementation and explains briefly the examples we used to
demonstrate how the aspects in Section \ref{sec:mool_problems} affect
the language. The code of our implementation, along with the examples,
is available at
\url{https://sourceforge.net/p/mool-plt-redex/code/ci/master/tree/mool1.rkt}.

Section \ref{sec:mool_revised} consists on our revision proposal for
the Mool language. We present a full formal system, consisting on the
revised operational semantics and a type system of the language, based
on the original but with changes that try to solve the aspects
identified in \ref{sec:mool_problems} plus the addition of new
features such as constructors.

Again, to test our revision we implemented the revised formalisation
using PLT-Redex. Section \ref{sec:mool_redex_2} presents the list of
examples used to test this second implementation. Most of these
examples are almost identical to the ones in section
\ref{sec:mool_redex_1}, but now they are expected to have a different
behaviour, while a few new examples that were used to test our changes
a little further. The code of the implementation, along with the
examples, is available at
\url{https://sourceforge.net/p/mool-plt-redex/code/ci/master/tree/mool2.rkt}.

\section{The original Mool language}
\label{sec:mool_problems}
Like said before, the main objective is to understand where Mool can
be too restrictive or even present incorrect behaviour. We did this by
not only reviewing the original
definitions~\cite{joanacorreiacampos2010}, but also by implementing
the language using PLT Redex and trying to falsify properties of the
system (see Section \ref{sec:mool_redex_1}). These aspects have been
categorised in major and minor aspects, based on their complexity.

\subsection{Minor errors and limitations}
\label{sec:mool_problems_minor}
The following observations are minor errors and limitations found on
Mool, i.e., they are very simple to solve:

\begin{enumerate}
\item The evaluation context for $\mkterm{while}$ is unnecessary. The
  evaluation contexts defined in the syntax of Mool specify that in a
  $\mkterm{while}$ expression the expression \textit{e} that serves as
  the boolean condition must be evaluated before the while expression
  itself, but the reduction rule \textsc{R-While} specifies that a
  $\mkterm{while}$ expression should be immediately reduced to a
  $\mkterm{if-else}$ expression.

\item \textsc{T-UsageVar} returns a new typing environment but it is
  not clear why the final environment needs to be be different from
  the initial.

\item \textsc{T-Assign} restricts assignments to unrestricted
  variables and fields only, but assignment to linear variables can be
  possible since any case that can risk linearity can be prevented by
  a predicate that checks if a variable has a linear type when it
  should not (for example, that already happens in rule
  \textsc{T-Class} where is specified that all of the class fields
  should be unrestricted).

%With the addition of local variables, since Mool has a method-level scope, in the end of the method evaluation the type checker should check if all variables declares inside the method are unrestricted in the same way the type checker checks if all of the class fields are unrestricted.

\item \textsc{T-Call} specifies that the parameter type should be the
  same as the method type, which is unnecessarily restricting.

\end{enumerate}

\subsection{Major errors and limitations}
\label{sec:mool_problems_major}
The following aspects are errors and limitations found on Mool that
are more complex to solve:
\begin{enumerate}
\item Subtyping for variant types is not well defined. The correct
  definition, based on the sub-typing definition in \cite{modular}, is
  as follows:

%\hspace*{\fill}If $\langle u' + u''\rangle <: u$ then $u = \langle u_t + u_f\rangle$ with $u_f <: u'$ and $u'' <:u$ \hspace*{\fill}

\begin{center}
  If $\langle u' + u''\rangle <: u$ then
  $u = \langle u_t + u_f\rangle$ with $u' <: u_t$ and $u'' <: u_f$
\end{center}

\item Subtyping seems to be unsafe. Consider the following expression:

\begin{center}
  $\mkterm{if}(f.eof()) \ \{ \ f.close(); \ false; \ \} \ \mkterm{else} \ \{ \ f.read(); \ true; \ \}$
\end{center}

In this expression, is the file has been fully read then it closes and returns $false$, informing the client that there is no more lines to read, otherwise it reads a line and returns true, informing the client that there is still lines to be read. Assume that we reverse the result output as follow:

\begin{center}
  $\mkterm{if}(f.eof()) \ \{ \ f.close(); \ true; \ \} \ \mkterm{else} \ \{ \ f.read(); \ false; \ \}$
\end{center}

Mool accepts this, but it can cause a runtime error because the client can try to close an already closed file. In this revision we will not propose a fix for the subtyping since it is not in the context of our work.

\item The typing rule \textsc{T-Spawn} states that the expression
  \textit{e} should have an unrestricted type, but that is not enough
  to prevent situations where the occurrence of statements being
  executed in different threads can result in the correct execution
  flow of a program being disrespected. For example, assuming a $File$
  class with the usage

%\hspace*{\fill}$\mkterm{lin} \{ open :  \mkterm{lin} \{ read :  \mkterm{lin} \{ close :  \mkterm{un}\{\} \} \} \}$ \hspace*{\fill}

\begin{center}
  $\mkterm{lin} \{ \ open : \mkterm{lin} \{ \ read : \mkterm{lin} \{ \ close
  : \ \mkterm{un}\{ \ \} \ \} \ \} \ \}$
\end{center}

where methods \textit{open}, \textit{read} and \textit{close} are all
of type $\mkterm{unit}$, the code

%\hspace*{\fill}$f.open(); \mkterm{spawn} \ f.read(); f.close()$ \hspace*{\fill}

\begin{center}
  $f.open(); \ \mkterm{spawn} \ f.read(); \ f.close()$
\end{center}

which opens the file, creates a separate thread for the reading
operation and closes the file, is wrong because after creating the new
thread with the reading operation it is not possible to predict the
next step, so the file can be read or closed. As defined, the type
system will accept this because $\ f.read()$ has type $\mkterm{unit}$,
which is an unrestricted type, and so the typing rule \textsc{T-Spawn}
will accept this expression, as the following partial derivation
shows:
\begin{mathpar}
 \inferrule*[right=T-Seq]
  {\inferrule*[right=T-Call]
   {\ldots}
  {\typedexp{\Tenv}{f.open()}{unit}{\Tenv'}} 
   \ \ \ \text{T1} }
  {\typedexp{\Tenv}{f.open(); \ \mkterm{spawn} \ f.read(); \ f.close()}{unit}{\Tenv'''}}
  
  \inferrule*[left = T1 \ , right=T-Seq]
  {
  \inferrule*[right=T-Spawn]
   {
	\inferrule*[right=T-Call]
   {\ldots}
  {\typedexp{\Tenv'}{f.read()}{unit}{\Tenv''}}    
   }
  {\typedexp{\Tenv'}{\mkterm{spawn} \ f.read()}{unit}{\Tenv''}} 
   \ \ \ \text{T2}
   }
  {\typedexp{\Tenv'}{\mkterm{spawn} \ f.read(); \ f.close()}{unit}{\Tenv''}}
  
  \inferrule*[left = T2 \ , right=T-Call]
   {\ldots}
  {\typedexp{\Tenv''}{f.close()}{unit}{\Tenv'''}} 
\end{mathpar}
  
  \begin{flushleft}
  $\Gamma = f : File[\mkterm{lin}\{ \ open : \mkterm{lin} \{ \ read : \mkterm{lin} \{ \ close : \ \mkterm{un} \ \{ \ \} \ \} \ \} \ \}]$\\
  $\Gamma' = f : File[\mkterm{lin} \{ \ read : \mkterm{lin} \{ \ close : \mkterm{un} \{ \ \} \ \} \ \}]$\\
  $\Gamma'' = f : File[close : \ \mkterm{un}\{ \ \}]$\\
  $\Gamma''' = f : File[\mkterm{un}\{ \ \}]$
  \end{flushleft}

\item Usages allow incorrect specifications of sequence of methods calls. Consider the following usage type:

\begin{center}
  $\mkterm{lin} \{ \ read : \rectype{Read}{ \mkterm{un} \{ \ eof : \langle
    close : \mkterm{un} \{ \} + read : Read \rangle \ \} } \ \} \ \}$
\end{center}

This usage describes a behaviour for a $File$ class, where method $read$ depends on variables initialized by a method $open$ that is implemented as a private method and is never called, that allows to read a line from a file before opening it but the typechecker allows it.

\item The type checker does not check if a field is initialised or
  not, allowing these to be dereferenced even when they are not.
  
\item The type system does not have typing rules for self
  calls. Although the typing rules for self calls were deliberately
  omitted from \cite{joanacorreiacampos2010}, they are essential since
  in case of recursion, the type system will not terminate the program
  evaluation. For instance, the method \textit{run} of the class
  $Seller$ of the example presented in Chapter 2 of
  \cite{joanacorreiacampos2010} is an example of a program that
  contains a self call that causes the type checker to go into an
  infinite loop.

\item Private methods are not evaluated since the type system, as
  defined, only checks methods in the class usage, which the system
  description considers public, and self calls are not included in the
  type system.

\item Typing rules for the control flow expressions with method calls
  as conditions are not applied when the method call is preceded of a
  negation, like

%\hspace*{\fill}$\mkterm{if}(!f.eof()) \ f.read() \ \mkterm{else} \ f.close() $ \hspace*{\fill}

\begin{center}
  $\mkterm{if}(!f.eof()) \ \{ \ f.read() \ \} \ \mkterm{else} \ \{ \ f.close() \ \} $
\end{center}

, treating these calls as regular expressions and so it does not
operate the necessary usage changes.

\item The language formalisation does not allow unrestricted classes,
  i.e., classes without usages.

\item $\mkterm{null}$ cannot be used as a value, not allowing the programmer to set objects to $\mkterm{null}$ or check if they are $\mkterm{null}$.
%\item A usage can go from unrestricted to linear, which goes against the system description.
%
%\hspace*{\fill}$\mkterm{lin} \{ open : \rectype{Read}{ \mkterm{un} \{eof : \langle close : \mkterm{un} \{ \} + read : Read \rangle \} } \}$ \hspace*{\fill}
%
%
%This usage, presented in the configuration of the core language, is a slightly modified usage to the File class of the example presented in \cite{joanacorreiacampos2010}.  The type system, as defined, will accept this usage but it clearly represents a situation where the usage goes from unrestricted to linear since when executing the method \textit{open} the usage goes from linear to unrestricted and when executing the method \textit{eof} the usage goes back to being linear.
%\end{enumerate}

\item An usage can go from an unrestricted state into a different
  state. According to the system description, an usage cannot go from
  an unrestricted state into a linear
  state.%, once it does into an unrestricted state, can go to other state.% can go from unrestricted to linear, which goes against the system description.

%\hspace*{\fill}$\mkterm{lin} \{ open : \rectype{Read}{ \mkterm{un} \{eof : \langle close : \mkterm{un} \{ \} + read : Read \rangle \} } \}$ \hspace*{\fill}
\begin{center}
  $\mkterm{lin} \{ \ open : \rectype{Read}{ \mkterm{un} \{ \ eof : \langle
    close : \mkterm{un} \{ \ \} + read : Read \rangle \ \} } \ \}$
\end{center}
This usage, presented in the configuration of the core language, is a
slightly modified usage to the File class of the example presented in
\cite{joanacorreiacampos2010}.  The type system, as defined, will
accept this usage but it clearly represents a situation where the
usage goes from unrestricted to linear since when executing the method
\textit{open} the usage goes from linear to unrestricted and when
executing the method \textit{eof} the usage goes back to being linear.

Although, the same concerns are valid when an usage is composed by
several unrestricted states and it transits between unrestricted
states. Consider a variation of the FileReader class that hosts a file
whose reading access can be blocked or unblocked. A possible usage
would be:
%
%\hspace*{\fill}$\mkterm{lin} \{ open : \rectype{Read}{ \mkterm{un} \{eof : \langle close : \mkterm{un} \{ \} + read : Read \rangle \} } \}$ \hspace*{\fill}

%\hspace*{\fill}$\mkterm{lin} \{ open : \rectype{Read}{ \mkterm{un} \{read; \rectype{Close}{ \mkterm{un} \{close : Done \} } \} } \}$ \hspace*{\fill}
%
%This usage, which is a variation of the $File$ usage, is composed by two unrestricted states, $Read$ and $Close$. This usage allows two clients to share the same file but it also allows both clients to start a reading operation but at the same time allowing one client to finish reading and closing the file while the other has not finished reading it.
\begin{center}
  $\mkterm{lin} \{ \ open : \mu{Blocked}.\mkterm{un} \{ \ unblock :$
  $ \mu{Unblocked}.\mkterm{un} \{ \ block; Blocked + read : Unblocked \ \}
  \ \} \ \}$
\end{center}

Consider also a situation where an instance of this $FileReader$
class, in state $Unblocked$, is shared between two clients. Since the
usage allows concurrent interaction with the instance, it is possible
for one client to execute $read$ and the other client to execute
$block$ at the same time and the $block$ operation terminates before
the $read$ operation. The client that is trying to read will do it
while the usage is in state $Blocked$, which is not the expected
behaviour. 

When in an unrestricted state, not only it must no return to a linear state it also must only go to the same state or to an equivalent state (i.e., a state with the exact same actions), like the following example:
%We consider that situations like these must not be prevented by the usage but by the programmer who has the responsibility to specify in which situations a method must be executed in mutual exclusion by a thread. Since Mool already provides the means for this, we will not go further in this context.
\begin{center}
  $\mkterm{lin} \{ \ open : \mu{Blocked}.\mkterm{un} \{ \ push :$
  $ \mu{Unblocked}.\mkterm{un} \{ \ push : Blocked \ \}
  \ \} \ \}$
\end{center}

\end{enumerate}

Although the original definition \cite{joanacorreiacampos2010} lacked
the ability to declare local variables, it was mentioned that the
implementation of Mool at the time had allowed it, so this aspect was
omitted from this list.

\section{Latest Mool implementation}
\label{sec:mool_compiler}

The work developed and presented in the following sections is based on
the Mool language presented in \cite{joanacorreiacampos2010}, but we
also reviewed the current Mool implementation available %
\footnote{The latest Mool implementation is available at
  \url{gloss.di.fc.ul.pt/tryit/Mool}}
to check if the aspects noted in Sections
\ref{sec:mool_problems_minor} and \ref{sec:mool_problems_major} still
remain or not and try to understand how the language copes with those
aspects. The examples used in this section are based on the
$FileAll.mool$ example.

To check if the subtyping in the current version is still unsafe, consider the following code:

\lstset{language=Java, label=lst:FileReader example 1}
\begin{lstlisting}[caption={$FileReader$ subtyping example},label={}]
			if(f.eof()) {
					f.close();        
					true;
			} else {
					s = s ++ f.read();        
					false;
			} 
\end{lstlisting}

While using this code as the body of the method $next$ of the $FileReader$ class, the compiler accepts it but running it will cause an infinite loop, which not only is a runtime error, it goes against the behaviour specified by the usage since the interaction with the file should be terminated after closing it, but in this example the $FileReader$ will execute the methods $eof$ and $close$. This proves that subtyping is still unsafe.

The compiler for the current Mool implementation checks if all of
class fields are initialised, even if they are not used, instead of
waiting for a runtime error, showing that the problem presented in
item 6 of Section \ref{sec:mool_problems_major} seems to be fixed. The
compiler also allows to assign values of linear type to variables,
showing that the restriction mentioned in item 3 of Section
\ref{sec:mool_problems_minor} was dropped, allowing code like this:

\lstset{language=Java, label=lst:FileReader example 1}
\begin{lstlisting}[caption={$FileReader$ linear attribution example},label={}]
			FileReader f; f = new FileReader();        
			f.open();        
    
			FileReader f2; f2 = new FileReader();
			f2.open();
    
			f = f2;
\end{lstlisting}
  
Moreover, it is possible to observe two aspects of the $\mkterm{spawn}$
construct: Mool does not allow \textit{e} to be a sequential
composition (it must only be a single expression) and not only it must
be a method call, it must consume that variable's usage. This last
aspect hints that the rule \textsc{T-Spawn} checks if all variables in
the typing environment are unrestricted after executing
\textit{e}. Using the example presented in item 2 of Section
\ref{sec:mool_problems_major}, with a class $File$ with the following
usage:%when that expression is a method call on a variable that variable's usage must be consumed by the end of that method call. This last aspect hints that the rule \textsc{T-Spawn} checks if all variables in the typing environment are unrestricted after executing \textit{e}. Using the example presented in item 2 of Section \ref{sec:mool_problems_major}, with a class File with the following usage:

\lstset{language=Java, label=lst:FileReader example 2}
\begin{lstlisting}[caption={$File$ usage variation},label={}]
			class File {   
					usage lin{open; Read} where
					    Read = lin{read ; Close}
					    Close = lin{close ; end};
					...                
			}
\end{lstlisting}

The following code, which is identical to the one from the example,
will not compile, with the compiler saying that it expected \textit{f}
to be \textit{null} in the third line :

\lstset{language=Java, label=lst:FileReader example 3}
\begin{lstlisting}[caption={$FileReader$ spawn example 1},label={}]
			File f; f = new File();
			f.open();
			spawn f.read();
			f.close();
\end{lstlisting}

However, the following code will compile, because the method \textit{close}
finalises the consumption of \textit{f}'s usage:

\lstset{language=Java, label=lst:FileReader example 4}
\begin{lstlisting}[caption={$FileReader$ spawn example 2},label={}]
			File f; f = new File();
			f.open();
			f.read();
			spawn f.close();
\end{lstlisting}

About the unsafe sequence of calls in item 4 of section \ref{sec:mool_problems_major}, consider the following example:

\lstset{language=Java, label=lst:listing_6}
\begin{lstlisting}[caption={$File$ unsafe usage},label={}]
			class File {   
					usage lin{read; Read} where 
						Read = lin{eof; 
						<lin{close; end} + lin{read; Read}>};               
			}
\end{lstlisting}

Replacing the original usage of the $FileAll.mool$ example with the one presented above will result in the program entering an infinite loop, due to the fact that the method $open$ is never called, meaning that both variables $linesRead$ and $linesInFile$ are never explicitly initialized and so both are initialized with the default value which is 0. It is valid to assume that, while the current version of Mool checks if a variable is initialized in the code, it seems to not check if that initialization happens during the execution of the program, leading to these type of situations.

About the use of negated calls as conditions in control flow
expressions, the current compiler still has this limitation. The
following example will not compile, saying that the method $read$ must
be called on a control flow expression:

\lstset{language=Java, label=lst:FileReader example 4}
\begin{lstlisting}[caption={$FileReader$ negated call example 1},label={}]
			if(!f.eof()) {
					s = s ++ f.read();        
					true;
			} else {
					f.close();        
					false;
			}
\end{lstlisting}

The message given by the compiler is not very clear since the method
$read$ is being called inside a control flow expression but the reason
for this error is due to the fact that, during the type-checking process, the
rule \textsc{T-If} is applied instead of the rule \textsc{T-IfV},
and it does not operate the necessary changes to the usage of the
field $f$ so that method $read$ is available to be called inside the
first branch and the method $close$ inside the second. Another example
is the following code where a $\mkterm{while}$ expression is used but
the compiler does not accept the code for the same reason as the
previous example:

\lstset{language=Java, label=lst:FileReader example 4}
\begin{lstlisting}[caption={$FileReader$ negated call example 2},label={}]
			while(!f.eof()) {
					s = f.read() ++ s;
			}
\end{lstlisting}

The current compiler allows classes to be unrestricted, as shown by
the example $PetitionAll.mool$ which has unrestricted classes such as
$Main$ and $PetitionServer$.

Furthermore, the current compiler does not allow an usage to go from
unrestricted to linear. The following example will not compile:

\lstset{language=Java, label=lst:listing_5}
\begin{lstlisting}[caption={$FileReader$ bad usage example 1},label={}]
			class File {   
					usage lin{open; Read} where
					    Read = un{eof; 
            				    <lin{close; end} + lin{read; Read}>};
					...                
			}
\end{lstlisting}

Furthermore, the current compiler does not allow an usage to go from
unrestricted to linear. The following example will not compile:

\lstset{language=Java, label=lst:listing_5}
\begin{lstlisting}[caption={$FileReader$ bad usage example 2},label={}]
			class File {   
					usage lin{open; Read} where
					    Read = un{eof; 
            				    <lin{close; end} + lin{read; Read}>};
					...                
			}
\end{lstlisting}

But the compiler can accept an usage that goes from an unrestricted state to another different unrestricted state, like the following one:

\lstset{language=Java, label=lst:listing_6}
\begin{lstlisting}[caption={$FileReader$ bad usage example 3},label={}]
			class FileReader {   
					usage lin{open; Blocked} where 
					    Blocked = un{unblock; Unblocked}
					    Unblocked = un{read; Unblocked + block; Blocked};
					...                
			}
\end{lstlisting}

\section{PLT Redex implementation of the original formalization}
\label{sec:mool_redex_1}
We implemented Mool as presented in \cite{joanacorreiacampos2010} using PLT Redex \footnote{Available at \ \url{https://sourceforge.net/p/mool-plt-redex/code/ci/master/tree/mool1.rkt}}. Due to the syntax of Racket, we had to make some modifications on the syntax of Mool, such as:
\begin{itemize}
\item Every expression must be in parenthesis.
\item \textbf{;} is reserved by Racket, so it cannot be used to separate expressions.
\item \textbf{.} is also reserved by Racket, so it was replaced by \textbf{->}.
\item To help implementing the type system, the usage variables \textbf{X} were replaced by \textbf{!X} so they could be distinguished from regular variables.
\item A new construct, $\mkterm{getref}$, was added to the runtime syntax. This new construct returns the last object identifier created so it can be assigned to a field.
\item In the runtime syntax used by the type system, nonterminals \textit{u} and \textit{D} were added to \textit{e} since there must be only one domain which, in this case, is \textit{e}.
\end{itemize}

In addition to the language implementation, the code also contains a few examples to show some of the problems noted in Section \ref{sec:mool_problems_major}. In order to implement more elaborate examples, some other changes were made:
\begin{itemize}
\item Items 1, 2 and 4 of Section \ref{sec:mool_problems_minor} are already solved in the implementation.
\item A typing rule for self calls was added. It is the same as \textsc{T-Call} but it does not change the usage, as the system description specifies.
\item Arithmetic and boolean expressions were implemented.
\end{itemize}

Finally, since $\mkterm{this}$ does not exist in this version, the object identifier 0 was reserved to represent $\mkterm{this}$, so every class field access and self call are done in 0. The examples are the following:
\begin{itemize}
\item[R-01] Implementation of the File example presented in \cite{joanacorreiacampos2010}, with an modification on how the program checks if it has reached the end of the file, due to the limitation presented in item 7 of section \ref{sec:mool_problems_major}. This example serves to test the operational semantics of Mool and when running it the reduction graph of the program's reduction will be shown.
\item[T-01] Typing example of the File example. When run the type system should be able to check the whole program with success.
\item[T-02] Typing example that implements the situation expressed in item 2 of Section \ref{sec:mool_problems_major}. The type checker verifies successfully when it should not.
\item[T-03] Implementation of the example presented in item 3 of Section \ref{sec:mool_problems_major}. The type checker evaluates the program successfully even though it is not desirable to have a situation where the file can be closed before being read.
\item[T-04] Same thing as T-01 but the fields $f$ from the $FileReader$ class and from the $Main$ class are not initialised, while both are dereferenced as in T-01. The program is evaluated successfully, allowing both fields to be dereferenced even though they are not initialised.
\item[T-05] Same thing as T-01 but in the usage of the $File$ class the method $open$ is replaced by the method $read$, same as the usage presented in item 4 of Section \ref{sec:mool_problems_major}. The usage allows to read the file without opening it but the typechecker verifies the program successfully.
\item[T-06] A variation of the $File$ example where the body of the method $count$ is changed to $true$. The return type of the method is $unit$ but the body of the method is of type $boolean$ and the type checker verifies the program successfully since the body of the method is not verified, only its signature;
\item[T-07] Implementation of the $File$ and $FileReader$ classes as presented in \cite{joanacorreiacampos2010}, including the using of a negated method call as a condition for a control flow expression in method $next$ of $FileReader$. This example serves to demonstrate the limitation presented in item 8 of Section \ref{sec:mool_problems_major} and it should fail.
\item[T-08] A variation of the $File$ and $FileReader$ classes, where now the method $next$ of $FileReader$ reads the whole file at once. This example is to demonstrate again the limitation presented in item 8 of section \ref{sec:mool_problems_major} with the same result, but now in a $\mkterm{while}$ expression.
\item[T-09] Another typing example that shows that the type system allows an usage to go from unrestricted to linear. This program only contains one class, $File$, but its usage is the same as the first example given in item 11 of Section \ref{sec:mool_problems_major}.
\item[T-10] A simplistic version of the $FileReader$ where the methods do not do anything but the usage, which is the same as the second usage presented in item 11 of Section \ref{sec:mool_problems_major}, is composed by two different unrestricted states and they change between them. This should not be allowed but the type checker allows it.
\end{itemize}

\section{The revised Mool language}
\label{sec:mool_revised}
This section presents our revision of the Mool language that tries to solve the problems mentioned in Sections \ref{sec:mool_problems_minor} and \ref{sec:mool_problems_major}. Some of the modifications are based on the observations made in Section \ref{sec:mool_compiler}.
\label{sec:mool_revised}

\subsection{Revised syntax}
\label{sec:mool_new_syntax}
Figure \ref{fig:mool_syntax} shows a modified syntax for the Mool
language. This revised syntax contains the following new/changed
elements:
\begin{enumerate}
\item Arithmetic and boolean expressions, represented by the
  nonterminals $a$ and $b$ respectively.
\item A new nonterminal $r$ for value references, which contains local
  variables \textit{d} and $\mkterm{this}$ (to help solving the
  problems noted in items 5 and 6 in Section
  \ref{sec:mool_problems_major}).
\item Expressions $e$ contain now only values and expressions,
  including calls, and put the rest of the constructs in a new
  nonterminal $s$ that represents statements.
\item Constructs $g \ d = e$ and $d = e$ to $s$ to allow local
  variable declaration and assignment.
\item Since we want to add the concept of constructor in the language,
  we modified the construct $\mkterm{new} \ C()$ to
  $\mkterm{new} \ C(e)$, allowing to pass parameters to the
  constructor.
\item We divided types into two nonterminals, $g$ and $t$. $g$ contains types
  that can be used to declare fields and variables, while $t$ contains every type
  in $g$ plus every other type such as $\mkterm{void}$ and $\nullterm$ 
\item We divided the usages into two nonterminals, $u$ and $z$. $u$ contains
  the usage constructs that can be used right at the beginning of the usage 
  while $z$ contains the usage constructs used during compile time. In the 
  runtime syntax we added $z$ to $u$ to avoid too many changes to the typing rules.
\item In the nonterminal \textit{u} we added $\epsilon$ to indicate
  that it is possible to not define an usage, making the class an
  unrestricted class.
\item The term \textit{o}, which are objects identifiers, is moved
  from the user syntax for the runtime syntax.
\item In the runtime syntax, a new type of value, $\mkterm{null}$, is
  added and it is used to represent values for non initialised
  objects, and a new type $\objecttype{C}{u;\vec{F}}$, where $\vec{F}$
  are mappings from fields that are initialised to types, is added to
  solve the problems in items 4 and 5 of Section
  \ref{sec:mool_problems_major}.
\item The evaluation context $\mkterm{while} \ \ectx \ e$ is removed
  for the reasons stated in item 1 of Section
  \ref{sec:mool_problems_minor}.
\end{enumerate}
%It is important that the type checker checks if a field is initialised when they are dereferenced to avoid the program to go into a error state. A solution proposal is to add a new runtime type, $\objecttype{C}{u;\vec{F}}$, where $\vec{F}$ are mappings from fields that are initialised to types, and modify the \textsc{T-Class} rule so instead of mapping all of the class fields to a type, it maps $\this$ to the type $\objecttype{C}{u;\vec{F}}$, where $\vec{F}$ will be empty. Every time a field is initialised, a new mapping is added to $\vec{F}$.

\begin{figure}
\input{mool_syntax}
  \caption{Revised syntax}
  \label{fig:mool_syntax}
\end{figure}

\subsection{Revised operational semantics}
\label{sec:mool_new_sos}
Figure \ref{fig:mool_reduc} shows the modified reduction rules for
this revised version of Mool. The rules differ from the original ones, as
%of the one used in \cite{joanacorreiacampos2010}, with the addition of
we add a new environment, \textit{local}, for the local variables.

We modified the rule \textsc{R-New} so that it reduces to a sequential
composition with the body of the constructor and the created object
identifier.

We also add the new rules \textsc{R-NewVar} and \textsc{R-AssignVar}
which are for local variable declaration and assignment.

The rule \textsc{R-AssignFieldNull}, allows to assign $\nullterm$
values to fields, removing them from the object's record.

Figures \ref{fig:arith-functions} and \ref{fig:bool-functions} show
the evaluation functions for the arithmetic and boolean
expressions. These functions, based on the ones presented in
\cite{riisnielsonh.nielsonf.2007}, receive as arguments an expression
and both the class field and local variable environment.

\begin{figure}
  \input{mool_semantics_operations}
  \caption{Auxiliary definitions and Operations}
  \label{fig:aux-defs-oper}
\end{figure}

\begin{figure}
  \centering
  \input{mool_semantics_states}
  \caption{Reduction semantics for states}
  \label{fig:mool_reduc}
\end{figure}

\begin{figure}
  \centering
  \input{mool_semantics_statements}
  \caption{Revised reduction semantics for statements}
  \label{fig:mool_reduc}
\end{figure}

\begin{figure}  
\input{mool_semantics_arith}
  \caption{Evaluation functions for arithmetic values and expressions}
  \label{fig:arith-functions}
\end{figure}

\begin{figure}  
\input{mool_semantics_bool}
  \caption{Evaluation functions for boolean values and expressions}
  \label{fig:bool-functions}
\end{figure}

\subsection{Revised type system}
\label{sec:mool_new_typing}
In this section we present a new set of typing rules. We omit the
unchanged rules with respect to the original
system~\cite{joanacorreiacampos2010}.

%Figure \ref{fig:mool-typing-programs} shows the proposed typing rules for programs, namely \textsc{T-Class}. This modified version of the rule has a new premise that checks if the class usage is correct, i.e., it does not go from unrestricted to linear at any point. Also, evaluation of the usage has an object $\objecttype{C}{u;\varnothing}$ for input, with no declared fields, and a object $\objecttype{C}{u';\vec{F}}$ for output, forcing the method-level scope of the system. In the the end it checks if all fields in $\vec{F}$ are unrestricted. 

Figure \ref{fig:mool-typing-programs} shows the proposed typing rules
for programs:
\begin{enumerate}
\item Rule \textsc{T-Class} is a modified version of the rule with the
  same name that has a new premise that checks if the class usage is
  correct, i.e., it does not go from an unrestricted state to a linear one at any point.

  Moreover, evaluation of the usage has an object
  $\objecttype{C}{u;\varnothing}$ for input, with no declared fields,
  and a object $\objecttype{C}{u';\vec{F}}$ for output, forcing the
  method-level scope of the system. In the end it checks if all fields
  in $\vec{F}$ are unrestricted.
\item Rule \textsc{T-UnClass} is for unrestricted classes and, instead
  of verifying the usage, it verifies all of the methods of the
  class.

  We assume that in unrestricted classes every method is independent,
  i.e., the changes it introduces to the state of the object do not
  affect other methods (e.g. initialised fields), so every method is
  verified using the same typing environments.
\end{enumerate}

Figure \ref{fig:mool-typing-usages} shows the proposed typing rules for usages:
\begin{enumerate}
\item Rule \textsc{T-BranchEnd} is a variation of \textsc{T-Branch}
  that is applicable when a usage branch terminates and so it does
  only evaluate the method, not the next usage (because there is
  none).
\item Rule \textsc{T-UsageVar} returns the same typing environment
  mapped to the usage variable, reflecting the observation made in the
  item 2 of \ref{sec:mool_problems_minor}. 
\end{enumerate}

Figures \ref{fig:mool-typing-arith} and \ref{fig:mool-typing-bool}
show the typing rules for the arithmetic and boolean
expressions. While the syntax itself already enforces the correct
types, we need these rules because the operands can change the usage,
e.g., a call made on a field as an operand.

Figure \ref{fig:mool-typing-deref} shows the proposed typing rules for field and variable dereference, where we added a new rule, \textsc{T-NullField}, for dereference of fields that have not initialized.

Figure \ref{fig:mool-typing-stms} shows the proposed typing rules for simple statements:
\begin{enumerate}
\item Two new rules, \textsc{T-NewVar} and \textsc{T-AssignVar}, for
  local variable declaration and assignment respectively, are added so
  the type checker can evaluate local variable declarations.

Both \textsc{T-AssignVar} and \textsc{T-AssignField} allow linear type
value assignment, solving the limitation in item 3 of Section
\ref{sec:mool_problems_minor}.
\item The rule \textsc{T-Spawn} is modified based on the conclusions
  presented in Section \ref{sec:mool_compiler}, making the type
  checker checking that:
  \begin{enumerate}
  \item all variables modified in $s$ are unrestricted; and
  \item in case of variables that are objects, the usage is fully
    consumed and therefore cannot be called in any other expression
    outside the $\mkterm{spawn}$;
  \item instead of checking if $s$ has unrestricted type, it allows
    expressions other than method calls and it also allows $s$ to be a
    sequential composition. 
  \end{enumerate}
\end{enumerate}

Figure \ref{fig:mool-typing-flow} shows the proposed typing rules for control flow expressions:
\begin{enumerate}
\item Rules \textsc{T-IfCall} and \textsc{T-WhileCall} are similar to
  the original rules \textsc{T-IfV} and \textsc{T-WhileV}, but we
  extended them so they can be applied to method calls made on local
  variables as the conditional expressions for these control flow
  expressions.

  Both these rules are replicated for unrestricted classes through
  rules \textsc{T-UnIfCall} and \textsc{T-UnWhileCall}, with the
  difference being that there is no usage modification because there
  is no usage, so they are essentially the rules \textsc{T-If} and
  \textsc{T-While} but instead of having a value as a condition they
  have a call on a object of a unrestricted class.
\item To solve the error in item 7 in section
  \ref{sec:mool_problems_major} we added the rules
  \textsc{T-IfNotCall} and \textsc{T-WhileNotCall}, which are similar
  to the rules \textsc{T-IfCall} and \textsc{T-WhileCall} but are for
  cases where the method call that serves as the condition is negated,
  resulting in the inverted attribution of the appropriate usage from
  the variant type given after the verification of the condition to
  the expressions that compose the control flow expression.
\item Rules \textsc{T-If} and \textsc{T-While}, which are for cases
  where the condition is simply a value and not a method call, are
  similar to the original rules with the same name, but the condition
  is a value \textit{v} instead of an expression \textit{e}.

  All four rules related to the $\mkterm{while}$ control
  flow expression were modified so that they allow modifications
  inside the loop, but to ensure that, in rules \textsc{T-WhileCall}
  and \textsc{T-WhileNotCall}, it is possible to execute the condition
  after executing the loop, both rules state that the type (and,
  consequently, the usage) of $w$ after the loop must be the same as
  the type $w$ has before executing the condition. 
\end{enumerate}

Figure \ref{fig:mool-typing-calls} shows the proposed typing rules for method calls:
\begin{enumerate}
\item We modified the rule \textsc{T-New} so that the constructor is
  evaluated has a call at the moment of initialisation and added the
  rule \textsc{T-UnNew} for unrestricted classes initialisation.
\item Rules \textsc{T-SelfCall1} and \textsc{T-SelfCall2}, which are
  for method calls made on $\mkterm{this}$, are added to solve the
  problem stated in item 5 in section \ref{sec:mool_problems_major}.

  Unlike the other typing rules for method calls, these do not change
  the usage, like the system description in
  \cite{joanacorreiacampos2010} specified, and check if the method was
  already evaluated or not, so that the type checker only checks a
  method body once in case of self calls, to prevent entering into a
  loop when the method is recursive. To check this, the
  \textit{methods} definition presented in \ref{fig:aux-defs-oper} is
  changed so that every method is associated to a boolean operator
  that informs if the method was already evaluated or not. This
  operator is ignored in the other method call rules.
\item Rule \textsc{T-Call} is similar to the original rule with the
  same name, but it is extended for method calls made on local
  variables and also with the minor error mentioned in item 4 of
  \ref{sec:mool_problems_minor} corrected.

  Moreover, due to the definition of the predicate $allows$, in
  particular the case when the usage is $\epsilon$, i.e., the class is
  unrestricted, the predicate also returns $\epsilon$, this rule can
  also be applied when the call is made on a object of an unrestricted
  class.
\end{enumerate}

About the subtyping not being safe, one possible solution would be modifying every $\mkterm{if}$-$\mkterm{else}$ typing rule to force both branches to be equivalent, i.e., to produce the same changes to the interacted objects. For example, consider the following derivation:

\begin{mathpar}
   \inferrule*[right=T-IfCall ]
   {\typedexp{\Tenv_1}{f.eof()}{\booltype}{\Tenv_2} \\ T1 \\ T2}
   {\typedexp{\Tenv_1}{\cond{f.eof()}{\{ \ f.close(); \ \mkterm{true}; \ \}}{ \{ \ f.read(); \ \mkterm{false}; \ \}}}{t}{\Tenv_6}} 
   
   \inferrule*[left= T1 \ , right = T-InjR ]
   { \inferrule*[right = T-Seq ]
   {\inferrule*[right = T-Call ]
   		{ }
   		{\typedexp{\Tenv_3}{f.close()}{\booltype}{\Tenv_5}} \\
   		\inferrule*[right = T-True ]
   			{ }
   			{\typedexp{\Tenv_5}{\mkterm{true}}{\booltype}{\Tenv_5}} }
   {\typedexp{\Tenv_3}{f.close(); \ \mkterm{true}}{\booltype}{\Tenv_5}}  }
   {\typedexp{\Tenv_3}{f.close(); \ \mkterm{true}}{\booltype}{\Tenv_6}} 
   
   \inferrule*[left= T2 \ , right = T-InjL ]
   { \inferrule*[right = T-Seq ]
   {\inferrule*[right = T-Call ]
   		{ }
   		{\typedexp{\Tenv_4}{f.read()}{\booltype}{\Tenv_1}} \\
   		\inferrule*[right = T-False ]
   			{ }
   			{\typedexp{\Tenv_1}{\mkterm{true}}{\booltype}{\Tenv_1}} }
   {\typedexp{\Tenv_4}{f.read(); \ \mkterm{false}}{\booltype}{\Tenv_1}}  }
   {\typedexp{\Tenv_4}{f.read(); \ \mkterm{false}}{\booltype}{\Tenv_6}} 
\end{mathpar}

  \begin{flushleft}
  $\Gamma_1 = f : File[Read]$\\
  $\Gamma_2 = f : File[\langle \mkterm{lin} \{ \ close :  \mkterm{un} \{ \ \} \ \} + \mkterm{lin} \{ \ read :  Read \ \} \rangle]$\\
  $\Gamma_3 = f : File[\mkterm{lin} \{ \ close :  \mkterm{un} \{ \ \} \ \}]$\\
  $\Gamma_4 = f : File[\mkterm{lin} \{ \ read :  Read \ \}]$\\
  $\Gamma_5 = f : File[un\{\ \}]$\\
  $\Gamma_6 = f : \langle \Gamma_1 + \Gamma_5 \rangle$
  \end{flushleft}
  
  This derivation is of the same example we used to show that subtyping can be unsafe in item 2 of section \ref{sec:mool_problems_major}, and demonstrates that the type system allows it to be verified. If we remove the subtyping this example would not pass, but neither would any other correct example. 
  
To show how the type system would behave without subtyping For example consider the following usage:
  \begin{mathpar}
  \mkterm{lin} \{ \ open : \rectype{Read}{ \mkterm{un} \{ \ eof : \langle
    close : \mkterm{un} \{ \ \} + read : \mkterm{un} \{ \ \} \rangle \ \} } \ \}
\end{mathpar}
  This usage is a variation of the $File$ usage, with the difference being that after executing $read$ the file is fully read and closes automatically. With this usage, the previous example would work without subtyping:
  
  \begin{mathpar}
   \inferrule*[right=T-IfCall ]
   {\typedexp{\Tenv_1}{f.eof()}{\booltype}{\Tenv_2} \\ T1 \\ T2}
   {\typedexp{\Tenv_1}{\cond{f.eof()}{ \{ \ f.close(); \ \mkterm{true}; \ \}}{\{ \ f.read(); \ \mkterm{false}; \ \}}}{t}{\Tenv_5}} 
   
   \inferrule*[left= T1 \ , right = T-Seq ]
   {\inferrule*[right = T-Call ]
   		{ }
   		{\typedexp{\Tenv_3}{f.close()}{\booltype}{\Tenv_5}} \\
   		\inferrule*[right = T-True ]
   			{ }
   			{\typedexp{\Tenv_5}{\mkterm{true}}{\booltype}{\Tenv_5}} }
   {\typedexp{\Tenv_3}{f.close(); \ \mkterm{true}}{\booltype}{\Tenv_5}}
   
   \inferrule*[left= T2 \ , right = T-Seq ]
   {\inferrule*[right = T-Call ]
   		{ }
   		{\typedexp{\Tenv_4}{f.read()}{\booltype}{\Tenv_1}} \\
   		\inferrule*[right = T-False ]
   			{ }
   			{\typedexp{\Tenv_1}{\mkterm{true}}{\booltype}{\Tenv_5}} }
   {\typedexp{\Tenv_4}{f.read(); \ \mkterm{false}}{\booltype}{\Tenv_5}} 
\end{mathpar}

  \begin{flushleft}
  $\Gamma_1 = f : File[Read]$\\
  $\Gamma_2 = f : File[\langle \mkterm{lin} \{ \ close :  \mkterm{un} \{ \ \} \ \} + \mkterm{lin} \{ \ read :  \mkterm{un} \{ \ \} \} \rangle]$\\
  $\Gamma_3 = f : File[\mkterm{lin} \{ \ close :  \mkterm{un} \{ \ \} \ \}]$\\
  $\Gamma_4 = f : File[\mkterm{lin} \{ \ read :  \mkterm{un} \{ \ \} \ \}]$\\
  $\Gamma_5 = f : File[un\{\ \}]$\\
  $\Gamma_6 = f : \langle \Gamma_1 + \Gamma_5 \rangle$
  \end{flushleft}
  
  Although this would work, it can be too restrictive to force both branches to leave the interacted object with the same usage in both returned environments. Because of this and the fact that proposing a more appropriate solution requires a deeper study that is out of the context of our work, we choose to ignore from now on.

\begin{figure}
  \input{mool_type_operations}
  \caption{Types, Type Definitions and Operations}
  \label{fig:types-defs}
\end{figure}

\begin{figure}
  \centering
  \input{mool_type_programs}
  \caption{Revised typing rules for programs}
  \label{fig:mool-typing-programs}
\end{figure}

\begin{figure}
  \centering
  \input{mool_type_usages}
  \caption{Revised typing rules for usages}
  \label{fig:mool-typing-usages}
\end{figure}

\begin{figure}
  \centering
  \input{mool_type_arith}
  \caption{Revised typing rules for arithmetic expressions}
  \label{fig:mool-typing-arith}
\end{figure}

\begin{figure}
  \centering
  \input{mool_type_bool}
  \caption{Revised typing rules for boolean expressions}
  \label{fig:mool-typing-bool}
\end{figure}

\begin{figure}
  \centering
  \input{mool_type_fields}
  \caption{Revised typing rules for field and variable dereference}
  \label{fig:mool-typing-deref}
\end{figure}

\begin{figure}
  \centering
  \input{mool_type_statements}
  \caption{Revised typing rules for simple statements}
  \label{fig:mool-typing-stms}
\end{figure}

\begin{figure}
  \centering
  \input{mool_type_control}
  \caption{Revised typing rules for control flow expressions}
  \label{fig:mool-typing-flow}
\end{figure}

\begin{figure}
  \centering
  \input{mool_type_calls}
  \caption{Revised typing rules for calls}
  \label{fig:mool-typing-calls}
\end{figure}

\section{PLT Redex implementation of the revised Mool formalization}
\label{sec:mool_redex_2}
To test our revision we implemented our formal system in PLT Redex.%
\footnote{Available at \
  \url{https://sourceforge.net/p/mool-plt-redex/code/ci/master/tree/mool2.rkt}}
Some of the examples in this version, aside from local variables and
the use of $\mkterm{this}$ as an value reference, are equal to the
ones in the PLT Redex implementation of Mool presented in Section
\ref{sec:mool_redex_1}. The examples presented in this version are the
following:
\begin{itemize}
\item[R-01] Implementation of the $File$ example presented in \cite{joanacorreiacampos2010} to test the operational semantics of Mool. Running it will result in the reduction graph of the program's reduction being shown.
\item[R-01] Implementation of the $File$ example presented in \cite{modular}.
%\item[R-01]  Implementation of the File example presented in \cite{joanacorreiacampos2010}.
\item[R-03] Example of a small program that uses an unrestricted class. The program contains the class $Folder$ which contains three methods independent from each other and a $Main$ class where a object of $Folder$ is created and interacted with. The $Main$ class could also be unrestricted but we defined it as linear to show the interaction of an unrestricted class through a linear one.
\item[R-04] Implementation of the $Auction$ example presented in \cite{joanacorreiacampos2010} that serves as a more complex test to the operational semantics of Mool.
\item[T-01] Typing example of the File example presented in \cite{joanacorreiacampos2010}. Should evaluate successfully.
\item[T-02] With the changes made to the \textsc{T-Spawn} rule, the type checker notices that executing the \textit{read} operation will modify the variable \textit{f} but will not consume its usage, which goes against  what is pretended, so the evaluation should fail.
\item[T-03] This program is similar to the one from T-05 but instead of creating a new thread for a reading operation, two separate threads are created for opening, reading and close separate files. This example, which evaluates successfully, shows that it is possible to use the construct $\mkterm{spawn}$ with several expressions.
\item[T-04] A similar example to T-01 but now the body of the method $main$ of class $Main$ is executed using $\mkterm{spawn}$. Although in the end the variable $f$ is unrestricted, it still can be used after the $\mkterm{spawn}$ expression, so it should fail because we changed the \textsc{T-Spawn} so that every usage modified inside a $\mkterm{spawn}$ expression should be at a $\mkterm{end}$, making it impossible to call any method from the object after the $\mkterm{spawn}$ expression.
\item[T-05] The field file of class $FileReader$ is not initialised but it is used, so the type checker will fail to evaluate because it checks if the field has already been initialised before using it.
\item[T-06] Same as T-01 but the constructor of the class $File$, where a number of variables are initialized, is replaced by the incrementation of the field $linesRead$, just like the method $read$. The typechecker does not accept this program;
\item[T-07] The type system now goes inside the body of private methods and verifies them, so in this example the verification will fail because the type checker notices that the return type of method $count$ is $void$ but the type of the body is $boolean$;
\item[T-08] Since now the type system is aware of which methods were already evaluated, this time the type checker will not enter in a infinite loop because it will only evaluate the the body of the recursive method \textit{read} of the class File once, ignoring its body when reaching the self call and thus evaluating the program successfully.
\item[T-09] Typing example of the File example presented in \cite{modular}. Should evaluate successfully.
\item[T-10] A variation of the File example where in the method $next$ of the $FileReader$ class, after closing the file the field $file$ is set to $\nullterm$. The type checker verifies the program successfully.
\item[T-11] With the new rule \textsc{T-Class} the type checker will detect that the usage goes from unrestricted to linear when executing the method \textit{eof}, so the evaluation should fail. 
\item[T-12] Again, the new rule \textsc{T-Class} also prevents an usage changing between different unrestricted states, so the program verification should fail;
\item[T-13] This example is similar to the one in T-12 but now the usage can change between equivalent unrestricted states.
\item[T-14] Typing example of the simple program introduced in R-03. Should evaluate successfully.
\item[T-15] Typing example of the $Auction$ example presented in \cite{joanacorreiacampos2010}. Should evaluate successfully.
\end{itemize}

\section{Conclusions and further work}
Following a detailed analysis of the formal definition and of the
implementation of the Mool programming language, we provide the
formalisation of a new version of the language with corrections of
errors and broader approaches to aspects where the language is too
restrictive. We also provide the implementation of the formalisation
of both the original and the revised versions using the Racket
programming language, more specifically its PLT Redex module, both
complemented with examples to help understanding the evolution between
versions.

The next stage of our work will be about the inference of usages from
programs written in a variation of Mool based on our revised
formalisation but it will not have usage annotations. Instead, the
programs will be equipped with assertions that we will use to infer
the usages.

%\bibliographystyle{abbrevnat}
%\bibliography{bibliography}
\printbibliography

\footnotesize
\newpage
\appendix
\section{Mool syntax} \label{App:AppendixA}

  % \centering
  \textbf{User Syntax}
  \begin{align*}
    \text{(class declarations)} && D& \bnf \class{C}{u}{\vec{F}}{\vec{M}}\\
    \text{(field declaration)} && F & \bnf g~f\\
    \text{(method declarations)} && M & \bnf y \ \methodd{t}{m}{t'\, x}{e} % \methoddd{t'}{m}{t\, x}{e}{p}{p'} 
    \\
	\text{(method qualifiers)} && y & \bnf \epsilon \alt \mkterm{sync} 
    \\\\
    \text{(values)} && v & \bnf \unit \alt n \alt \true\alt\false \alt \nullterm\\
    \text{(local value references)}                        && r & \bnf d \alt \mkterm{this} \\
	\text{(global value references)}                        && w & \bnf r \alt r.f \\    
    \text{(calls)} && c & \bnf \mkterm{new} \ C (e) \alt \methcal{r}{m}{e} \alt r.f.m(e)\\    
    \text{(arithmetic operations)} && a & \bnf n \alt w \alt c \alt a + a \alt a - a \\ &&& \quad\alt a * a \alt a / a\\    
	\text{(boolean operations)} && b & \bnf \true \alt \false \alt w \alt c \alt a == a \alt a \ != a \\ &&& \quad\alt a <= a \alt a >= a \alt a < a \alt a > a \\ &&& \quad\alt b \ \&\& \ b \alt b \ || \ b \alt !b\\
    \text{(expressions)} && e & \bnf v \alt a \alt b \\ &&& \quad\alt c \alt w\\
    \text{(statements)}  && s & \bnf e \alt \seq{s}{s'} \\ &&& \quad\alt \upd{r}{f}{e} \alt g \ d = e \alt d = e \\
   %&&& \quad\alt \\
   &&& \quad\alt \cond{b}{s'}{s''} \alt \while{b}{s'} \\
   &&& \quad\alt \mkterm{spawn}\{s\}\\
   \text{(types)} &&   t & \bnf \unittype\alt g \alt\nullterm\\
   \text{(declarable types)} &&   g & \bnf \mkterm{int}\alt\booltype\alt\objecttype{C}{z}\\
   \text{(class usages)} && u & \bnf \epsilon \alt q\echoicetypel{m}{z}{i}{I} \alt \rectype{X}{u}\\
   \text{(usages)} && z & \bnf u  \alt \langle u + u \rangle \alt X\\   
   \text{(usage types)} && q & \bnf \unterm \alt \linterm \\\\
  \end{align*}  
  \textbf{Runtime Syntax}  
  \begin{align*}
    %\text{(values)} && v & \bnf \ldots\alt \nullterm \alt o\\
    \text{(values)} && v & \bnf \ldots \alt o\\
    \text{(value references)} && r & \bnf \ldots\alt o\\
    \text{(class usages)} && u & \bnf \ldots \alt z\\   
    \text{(types)} && t & \bnf \ldots\alt \objecttype{C}{z;\vec{F}}\\
    \\
    \text{(object records)} && R & \bnf ( C,u,\overrightarrow{f = v},l)\\
    \text{(field value map)} && l & \bnf 0 \alt 1\\
    \text{(heap)} && h & \bnf \emptyset \alt h, o = R\\
    \text{(evaluation context)} && \ectx & \bnf [-] \alt \seq{\ectx}{s} \alt \upd{o}{f}{\ectx} \alt \selfcal{\der{o}{m}}{\ectx} \alt \methcal{\der{o}{f}}{m}{\ectx} \\
    &&&\quad \alt \cond{\ectx}{s}{s'} \\
    \text{(States)} && S & \bnf \stat{h, local}{s_1 \alt \ldots \alt s_n}
  \end{align*}
  
\newpage
\section{Mool revised operational semantics}

\subsection{Auxiliary definitions and Operations}
\textbf{Object Record and Heap Operations}
  \begin{align*}
    \langle C,u,\vec{V}\rangle.f & \defeq \vec{V}(f) &
    \langle C,u,\vec{V}\rangle.\usageterm & \defeq u\\
    \langle C,u,\vec{V}\rangle.\classterm & \defeq C  \\
    \end{align*}
    \textbf{Operations for values and types}
    \begin{align*}
	\linterm(v) & \defeq
                       \begin{cases}
                         tt  & \text{ if } v = o \wedge h(v).\usageterm = \langle u' + u'' \rangle\\
                         tt  & \text{ if } v = o \wedge h(v).\usageterm = \linterm\echoicetypel{m}{z}{i}{I}\\
                         \mathit{ff} &	\text{ otherwise }
                       \end{cases} &
    \unterm(v) & \defeq
                       \begin{cases}
                         tt  & \text{ if } v = \mkterm{unit}\\
                         tt  & \text{ if } v = n\\
                         tt  & \text{ if } v = \true\\
                         tt  & \text{ if } v = \false\\
                         tt  & \text{ if } v = o \wedge h(v).\usageterm = \unterm\echoicetypel{m}{z}{i}{I}\\
                         \mathit{ff} &	\text{ otherwise }
                       \end{cases}
    \end{align*}
   \textbf{Class Definition Operations}
    \begin{align*} 
    C.\methsterm  & \defeq \overrightarrow{M,eval} \quad\text{ where }\class{C}{u}{\vec{F}}{\vec{M}} \in \vec{D} \ \text{and} \ eval \in \{ 0, 1 \}  \\
    C.\fieldsterm  & \defeq \vec{F} \quad\text{ where }\class{C}{u}{\vec{F}}{\vec{M}} \in \vec{D} \\
    C.\usageterm  & \defeq u  \quad\text{ where }\class{C}{u}{\vec{F}}{\vec{M}} \in \vec{D} 
  \end{align*}

\subsection{Reduction semantics for states}  
\begin{mathpar}
  \inferrule*[left=R-Context]
  {\stat{h, local}{s_1 \alt \ldots \alt s \alt \ldots \alt s_n}\reduces \stat{h', local'}{s_1 \alt \ldots \alt s' \alt \ldots \alt s_n}}
  {\stat{h, local}{s_1 \alt \ldots \alt \ectx[s] \alt \ldots \alt s_n}\reduces \stat{h', local'}{s_1 \alt \ldots \alt \ectx[s'] \alt \ldots \alt s_n}}
  \and
  \inferrule*[left=R-Spawn]
  {}
  {\stat{h, local}{s_1 \alt \ldots \alt \ectx[\mkterm{spawn}\{s\}] \alt \ldots \alt s_n}\reduces \stat{h, local}{s_1 \alt \ldots \alt \ectx[\mkterm{unit}] \alt \ldots \alt s_n}}
  \end{mathpar}
    
\subsection{Evaluation functions for arithmetic values and expressions}
\begin{mathpar}
\mathcal{N} (n)  = n \ \ \ \ \ \ \ \
	\mathcal{A}(n, h, local)  = \mathcal{N}(n) \\
	\mathcal{A}(o.f, h, local)  = h(o).f \ \ \ \ \ \ \ \
	\mathcal{A}(d , h, local)  = local(d) \\
	\mathcal{A}(a_1 + a_2, h, local)  = \mathcal{A} (a_1, h, local) + \mathcal{A} (a_2, h, local) \\
	\mathcal{A}(a_1 - a_2, h, local)  = \mathcal{A} (a_1, h, local) - \mathcal{A} (a_2, h, local)\\
	\mathcal{A}(a_1 * a_2, h, local)  = \mathcal{A} (a_1, h, local) * \mathcal{A} (a_2, h, local) \\ 
	\mathcal{A}(a_1 \ / \ a_2, h, local)  = \mathcal{A} (a_1, h, local) \ / \ \mathcal{A} (a_2, h, local)
\end{mathpar}

\subsection{Evaluation functions for boolean values and expressions}
\begin{mathpar}
\mathcal{B}(\true, h, local)  = \true \ \ \ \ \ \ \ \ \mathcal{B}(\false, h, local)  = \false \\ 
\mathcal{B}(o.f, h, local)  = h(o).f  \ \ \ \ \ \ \ \ \mathcal{B}(d , h, local)  = local(d) \\
\mathcal{B}(a_1 == a_2 , h, local) = 
	\begin{cases} 
		\true & \mathcal{A} (a_1, h, local) = \mathcal{A} (a_2, h, local) \\
		\false & \mathcal{A} (a_1, h, local) \neq \mathcal{A} (a_2, h, local)		
	\end{cases} \\
\mathcal{B}(a_1 != a_2 , h, local) = 
	\begin{cases} 
		\true & \mathcal{A} (a_1, h, local) \neq \mathcal{A} (a_2, h, local) \\
		\false & \mathcal{A} (a_1, h, local) = \mathcal{A} (a_2, h, local)		
	\end{cases} \\
\mathcal{B}(a_1 < a_2 , h, local) = 
	\begin{cases} 
		\true & \mathcal{A} (a_1, h, local) < \mathcal{A} (a_2, h, local)\\
		\false & \mathcal{A} (a_1, h, local) >= \mathcal{A} (a_2, h, local)
	\end{cases} \\
\mathcal{B}(a_1 < a_2 , h, local) = 
	\begin{cases} 
		\true & \mathcal{A} (a_1, h, local) < \mathcal{A} (a_2, h, local)\\
		\false & \mathcal{A} (a_1, h, local) <= \mathcal{A} (a_2, h, local)
	\end{cases} \\
\mathcal{B}(a_1 <= a_2 , h, local) = 
	\begin{cases} 
		\true & \mathcal{A} (a_1, h, local) <= \mathcal{A} (a_2, h, local)\\
		\false & \mathcal{A} (a_1, h, local) > \mathcal{A} (a_2, h, local)
	\end{cases} \\
\mathcal{B}(a_1 >= a_2 , h, local) = 
	\begin{cases} 
		\true & \mathcal{A} (a_1, h, local) >= \mathcal{A} (a_2, h, local)\\
		\false & \mathcal{A} (a_1, h, local) < \mathcal{A} (a_2, h, local)
	\end{cases} \\
\mathcal{B}(b_1 \ \&\& \ b_2 , h, local) = 
	\begin{cases} 
		\true & \mathcal{B} (b_1, h, local) = \true \wedge (b_2, h, local) = \true \\
		\false & \text{otherwise}
	\end{cases} \\
\mathcal{B}(b_1 \ || \ b_2 , h, local) = 
	\begin{cases} 
		\true & \mathcal{B} (b_1, h, local) = \true \vee (b_2, h, local) = \true \\
		\false & \text{otherwise}
	\end{cases} \\
\mathcal{B}(!b, h, local) = 
	\begin{cases} 
		\true & \mathcal{B} (b, h, local) = \false\\
		\false & \mathcal{B} (b, h, local) = \true
	\end{cases} \\
\end{mathpar}

\subsection{Revised reduction semantics for statements}  
\begin{mathpar}
    \inferrule*[left=R-UnField \ ]
    {h(o).f=v\\ \unterm(v,h)}
    {\stat{h,local}{\der{o}{f}} \reduces \stat{h,local}{v}}
    \and
   \inferrule*[left=R-LinField \ ]
   {h(o).f=v\\ \linterm(v,h)}
   {\stat{h,local}{\der{o}{f}} \reduces \stat{\changeval{h}{\der{o}{f}}{\nullterm},local}{v}}
   \and
   \inferrule*[left=R-UnVar \ ]
    {h(d)=v\\ \unterm(v,h)}
    {\stat{h,local}{d} \reduces \stat{h,local}{v}}
    \and
   \inferrule*[left=R-LinVar \ ]
   {h(d)=v\\ \linterm(v,h)}
   {\stat{h,local}{d} \reduces \stat{\changeval{h}{d}{\nullterm},local}{v}}
   \and
    \inferrule*[left=R-Seq \ ]
    {}
    {\stat{h,local}{\seq{v}{s}} \reduces \stat{h,local}{s}} 
    \and
    \inferrule*[left=R-NewVar \ ]
    {}
    {\stat{h,local}{g \ d = v } \reduces \stat{h,\changeval{local}{d}{v}}{\unit}}
    \and
    \inferrule*[left=R-AssignVar \ ]
    {}
    {\stat{h,local}{d = v} \reduces \stat{h,\changeval{local}{d}{v}}{\unit}}
    \and
    \inferrule*[left=R-AssignField \ ]
    {v \neq \nullterm}
    {\stat{h,local}{\upd{o}{f}{v}} \reduces \stat{\changeval{h}{\der{o}{f}}{v},local}{\unit}}
    \and
	 \inferrule*[left=R-AssignFieldNull \ ]
    {v = \nullterm}
    {\stat{h,local}{\upd{o}{f}{v}} \reduces \stat{h \setminus o.f,local}{\unit}}
    \and   
    \inferrule*[left=R-New \ ]
    {o \text{ fresh} \\ (\methodd{\_ }{C}{\_ ~ x}{s}, \_) \in C.\methsterm \\ C.\fieldsterm = \overrightarrow{t~f} \\ C.\usageterm = u }
    {\stat{h,local}{\mkterm{new} \ C(v)}\reduces \stat{(h,o=\langle C, u, \overrightarrow{f=\nullterm}\rangle),local}{s\subs{o}{\this}\subs{v}{x}; o}}
  \and
   \inferrule*[left=R-Call \ ]
   { (\methodd{\_ }{m}{\_ ~ x}{s}, \_) \in (h(o).\classterm).\methsterm  }
   {\stat{h,local}{\selfcal{\der{o}{m}}{v}} \reduces \stat{h,local}{s\subs{o}{\this}\subs{v}{x}}}
  \and
   \inferrule*[left=R-FieldCall \ ]
   { (\methodd{\_ }{m}{\_ ~ x}{s}, \_) \in (h(o).f.\classterm).\methsterm }
   {\stat{h,\varnothing}{\methcal{\der{o}{f}}{m}{v}} \reduces \stat{h,\varnothing}{s\subs{o}{\this}\subs{v}{x}}}
  \and
  \inferrule*[left=R-While \ ]
  {}
  {\stat{h,local}{\while{b}{s}}\reduces \stat{h,local}{\cond{b}{(\seq{s}{\while{b}{s}})}{\unit}}}
  \and
  \inferrule*[left=R-IfTrue \ ]
  {}
  {\stat{h,local}{\cond{\true}{s'}{s''}}\reduces \stat{h,local}{s'}}
 \and
  \inferrule*[left=R-IfFalse \ ]
  {}
  {\stat{h,local}{\cond{\false}{s'}{s''}}\reduces \stat{h,local}{s''}}
  \end{mathpar}
  
\newpage
\section{Mool revised type system}
\subsection{Type operations}
  \textbf{Type Operations}
  \begin{align*}
  \linterm(t) & \defeq
                       \begin{cases}
                         tt  & \text{ if } v = o \wedge h(v).\usageterm = \langle u' + u'' \rangle\\
                         tt  & \text{ if } v = o \wedge h(v).\usageterm = \linterm\echoicetypel{m}{z}{i}{I}\\
                         \mathit{ff} &	\text{ otherwise }
                       \end{cases}\\
    \unterm(t) & \defeq
                       \begin{cases}
                         tt  & \text{ if } v = \unittype\\
                         tt  & \text{ if } v = \mkterm{int}\\
                         tt  & \text{ if } v = \booltype\\
                         tt  & \text{ if } v = o \wedge h(v).\usageterm = \unterm\echoicetypel{m}{z}{i}{I}\\
                         \mathit{ff} &	\text{ otherwise }
                       \end{cases}  \\
       \linterm(\Tenv) & \defeq \forall \ (t \ f) \in \Tenv : \linterm(t)\\
    \unterm(\Tenv) & \defeq \forall \ (t \ f) \in \Tenv : \unterm(t)\\
    \linterm(\vec{F}) & \defeq \forall \ (t \ f) \in \vec{F} : \linterm(t)\\
    \unterm(\vec{F}) & \defeq \forall \ (t \ f) \in \vec{F} : \unterm(t)\\
   \mkterm{check}(\Phi, u) & \defeq
                       \begin{cases}
                       	  \mkterm{check}(\Phi, u_i) &  u = \linterm\echoicetypel{m}{u}{i}{I} \\
                       	  \mkterm{check}(\Phi, u_t) \wedge \mkterm{check}(\Theta, u_f) & u = \langle u_t + u_t \rangle \\
                          \mkterm{check}((\Phi, X : u),u)  & u = \rectype{X}{u} \\
                          \mathit{ff} & u = \unterm\echoicetypel{m}{u}{i}{I} \wedge \exists u_n \in u_i: u_n = \rectype{X}{u_X}\\ 
                          \mathit{ff} & u = \unterm\echoicetypel{m}{u}{i}{I} \wedge \exists u_n \in u_i: u = \linterm\echoicetypel{m}{u}{j}{J}\\ 
                          \mathit{ff} & u = \unterm\echoicetypel{m}{u}{i}{I} \wedge \exists u_n \in u_i: u_n = X \ \wedge \\ & \ \ \ \ \ \  \Phi(X) = \langle u_t + u_t \rangle\\  
                           \mathit{ff} & u = \unterm\echoicetypel{m}{u}{i}{I} \wedge \exists u_n \in u_i: u_n = X \ \wedge \\ & \ \ \ \ \ \  \Phi(X) = \linterm\echoicetypel{m}{u}{j}{J}\\                                  
                          \mathit{ff} & u = \unterm\echoicetypel{m}{u}{i}{I} \wedge \exists u_n \in u_i: u = \langle u_t + u_t \rangle\\                   
                          \mathit{ff} & u = \unterm\echoicetypel{m}{u}{i}{I} \wedge \exists u_n \in u_i: u = \unterm\echoicetypel{m}{u}{j}{J} \wedge m_i \neq m_j\\   
                          \mathit{ff} & u = \unterm\echoicetypel{m}{u}{i}{I} \wedge \exists u_n \in u_i: u_n = X \ \wedge \\ & \ \ \ \ \ \  \Phi(X) = \unterm\echoicetypel{m}{u}{j}{J} \wedge m_i \neq m_j\\                            
                          tt & \text{otherwise}
                       \end{cases}  \\        
    u.\allowsterm(m_j) & \defeq
                       \begin{cases}
                       \epsilon & u = \epsilon \\
                         \unterm  & u=\unterm \\
                         u_j  & u =  \echoicetypel{m}{u}{i}{I} \text{ and } j\in I\\
                         u'\subs{\rectype{X}{u'}}{X}.\allowsterm(m_j)               & u = \rectype{X}{u'}\\
                          \text{undefined} & \text{otherwise}
                       \end{cases}  \\
    \agreeterm(t,t') & \defeq
                       \begin{cases}
                         tt  & \text{ if } t=t'=\booltype\\
                         tt  & \text{ if } t=t'=\mkterm{int}\\
                         tt  & \text{ if } t=t'=\unittype\\
                         tt  & \text{ if } t = \objecttype{C}{u} \text{ and }  t' = \objecttype{C}{u;\vec{F}}\\
                         tt  & \text{ if } t = \mkterm{null} \text{ or }  t' = \mkterm{null}
                       \end{cases}  \\
    \mkterm{modified}(\Tenv, \Tenv') & \defeq
                      \forall r \in \Tenv': \ r \notin \Tenv \vee \Tenv(r) \neq \Tenv'(r)\\
    \mkterm{completed}(\Tenv, \Tenv') & \defeq
                      \forall r \in \Tenv': (r \notin \Tenv \vee \Tenv(r) \neq \Tenv'(r)) \wedge (r = \objecttype{C}{u;\vec{F}} \implies u = \mkterm{un}\{\})
\end{align*}

\subsection{Revised typing rules for programs}
\begin{mathpar}
   \inferrule*[left=T-Class]
   {\text{check}(\emptyset, u) \\ \objecttype{C}{u;\emptyset} \triangleright u \triangleleft \objecttype{C}{u;\vec{F}} \\ \text{un}(\vec{F})}
   {\typedclass{\class{C}{u}{\vec{F}}{\vec{M}}}}
   \and
   \inferrule*[left=T-UnClass]
   {\forall i\in I \cdot \\\\ %\methodd{\_ }{m}{\_ ~ x}{s}
    (\methodd{y \ t}{m_i}{t'\, x}{s}, \_) \in \vec{M}  \\  \objecttype{C}{\emptyset}, x : t' \triangleright s \triangleleft \objecttype{C}{\vec{F}} \\ \text{un}(\vec{F})}
   {\typedclass{\mkterm{class} \ C \{\vec{F};\vec{M}\}}}  
  \end{mathpar}
  
\subsection{Revised typing rules for usages}
\begin{mathpar}
  \iffalse
   \inferrule*[left=T-BranchEnd \ ]
    % \infrule
    {\forall i\in I \cdot \\\\    
        (\methodd{y \ t}{m}{t'\, x}{s}, \_) \in C.\methsterm  \\
        \typedexp{\this:\objecttype{C}{u;\vec{F}},x:t'}{s}{t}{\Tenv} \\
        \typedval{\Tenv}{\this}{\objecttype{C}{u_i;\vec{F_i}}} \\
    }
    {\Theta; \objecttype{C}{u;\vec{F}}  \triangleright \_ \{m ; un\{\}\} \triangleleft \ \Gamma}
    \fi
   \inferrule*[left=T-Branch \ ]
    % \infrule
    {\forall i\in I \cdot \\\\    
        (\methodd{y \ t}{m_i}{t'\, x}{s}, \_) \in C.\methsterm  \\
        \typedexp{\this:\objecttype{C}{u;\vec{F}},x:t'}{s}{t}{\Tenv} \\\\
        \typedval{\Tenv}{x}{t''} \\ \text{un}(t'') \\
        \typedval{\Tenv}{\this}{\objecttype{C}{u_i;\vec{F_i}}} \\
        \Theta; \Gamma  \triangleright u_i \triangleleft \ \Gamma' 
    }
    {\Theta; \objecttype{C}{u;\vec{F}}  \triangleright \_ \echoicetypel{m}{u}{i}{I} \triangleleft \ \Gamma'}
    \and
    \inferrule*[left=T-BranchEnd \ ]
    % \infrule
    {}
    {\Theta; \Gamma  \triangleright un\{\} \triangleleft \ \Gamma}
    \and
    \inferrule*[left=T-UsageVar \ ]
    % \infrule
    {}
    {(\Theta, X : \Gamma); \Gamma \triangleright X \triangleleft \ \Gamma}
  \end{mathpar}
\begin{mathpar}
    \inferrule*[left=T-Variant \ ]
    % \infrule
    {\Theta; \Gamma' \triangleright u_t \triangleleft \ \Gamma \\ \Theta; \Gamma'' \triangleright u_f \triangleleft \ \Gamma}
    {\Theta; \langle \Gamma' + \Gamma'' \rangle \triangleright \langle u_t + u_f \rangle \triangleleft \ \Gamma}
    \and
    \inferrule*[left=T-Rec \ ]
    % \infrule
    {(\Theta, X : \Gamma); \Gamma \triangleright u \triangleleft \ \Gamma'}
    {\Theta; \Gamma \triangleright \rectype{X}{u} \triangleleft \ \Gamma'}
  \end{mathpar}
  
\subsection{Revised typing rules for values}
\begin{mathpar}
  \inferrule*[left=T-Unit \ ]
  {}
  {\typedexp{\Tenv}{\mkterm{unit}}{\mkterm{void}}{\Tenv}}
  \and
  \inferrule*[left=T-Int \ ]
  {}
  {\typedexp{\Tenv}{n}{\mkterm{int}}{\Tenv}}
  \and
  \inferrule*[left=T-True \ ]
  {}
  {\typedexp{\Tenv}{\mkterm{true}}{\mkterm{boolean}}{\Tenv}}
  \and
  \inferrule*[left=T-False \ ]
  {}
  {\typedexp{\Tenv}{\mkterm{false}}{\mkterm{boolean}}{\Tenv}}
  \and
  \inferrule*[left=T-Null \ ]
  {}
  {\typedexp{\Tenv}{\mkterm{null}}{\mkterm{null}}{\Tenv}}
  \end{mathpar}
  
\subsection{Revised typing rules for arithmetic expressions}
\begin{mathpar}
    \inferrule*[left=T-Add \ ]
  {\typedexp{\Tenv}{a_1}{\mkterm{int}}{\Tenv'} \\ \typedexp{\Tenv'}{a_2}{\mkterm{int}}{\Tenv''}}
  {\typedexp{\Tenv}{a_1 + a_2}{\mkterm{int}}{\Tenv''}}
  \and
  \inferrule*[left=T-Sub \ ]
  {\typedexp{\Tenv}{a_1}{\mkterm{int}}{\Tenv'} \\ \typedexp{\Tenv'}{a_2}{\mkterm{int}}{\Tenv''}}
  {\typedexp{\Tenv}{a_1 - a_2}{\mkterm{int}}{\Tenv''}}
  \and
  \inferrule*[left=T-Mult \ ]
  {\typedexp{\Tenv}{a_1}{\mkterm{int}}{\Tenv'} \\ \typedexp{\Tenv'}{a_2}{\mkterm{int}}{\Tenv''}}
  {\typedexp{\Tenv}{a_1 * a_2}{\mkterm{int}}{\Tenv''}}
  \and
  \inferrule*[left=T-Div \ ]
  {\typedexp{\Tenv}{a_1}{\mkterm{int}}{\Tenv'} \\ \typedexp{\Tenv'}{a_2}{\mkterm{int}}{\Tenv''}}
  {\typedexp{\Tenv}{a_1 \ / \ a_2}{\mkterm{int}}{\Tenv''}}
  \end{mathpar}
  
\subsection{Revised typing rules for boolean expressions}
\begin{mathpar}
    \inferrule*[left=T-Eq \ ]
  {\typedexp{\Tenv}{e_1}{t}{\Tenv'} \\ \typedexp{\Tenv'}{e_2}{t'}{\Tenv''} \\ \agreeterm(t',t)}
  {\typedexp{\Tenv}{e_1 == e_2}{\mkterm{bool}}{\Tenv''}}
  \and  
  \inferrule*[left=T-Diff \ ]
  {\typedexp{\Tenv}{e_1}{t}{\Tenv'} \\ \typedexp{\Tenv'}{e_2}{t'}{\Tenv''} \\ \agreeterm(t',t)}
  {\typedexp{\Tenv}{e_1 != e_2}{\mkterm{bool}}{\Tenv''}}
  \and
  \inferrule*[left=T-Greater \ ]
  {\typedexp{\Tenv}{a_1}{\mkterm{int}}{\Tenv'} \\ \typedexp{\Tenv'}{a_2}{\mkterm{int}}{\Tenv''}}
  {\typedexp{\Tenv}{a_1 > a_2}{\mkterm{bool}}{\Tenv''}}
  \and
  \inferrule*[left=T-Less \ ]
  {\typedexp{\Tenv}{a_1}{\mkterm{int}}{\Tenv'} \\ \typedexp{\Tenv'}{a_2}{\mkterm{int}}{\Tenv''}}
  {\typedexp{\Tenv}{a_1 < a_2}{\mkterm{bool}}{\Tenv''}}
  \and
  \inferrule*[left=T-GtEqual \ ]
  {\typedexp{\Tenv}{a_1}{\mkterm{int}}{\Tenv'} \\ \typedexp{\Tenv'}{a_2}{\mkterm{int}}{\Tenv''}}
  {\typedexp{\Tenv}{a_1 >= a_2}{\mkterm{bool}}{\Tenv''}}
	\and
	\inferrule*[left=T-LeEqual \ ]
  {\typedexp{\Tenv}{a_1}{\mkterm{int}}{\Tenv'} \\ \typedexp{\Tenv'}{a_2}{\mkterm{int}}{\Tenv''}}
  {\typedexp{\Tenv}{a_1 <= a_2}{\mkterm{bool}}{\Tenv''}}
  \and
  \inferrule*[left=T-And \ ]
  {\typedexp{\Tenv}{b_1}{\mkterm{bool}}{\Tenv'} \\ \typedexp{\Tenv'}{b_2}{\mkterm{bool}}{\Tenv''}}
  {\typedexp{\Tenv}{b_1 \ \&\& \ b_2}{\mkterm{bool}}{\Tenv''}}
  \and
   \inferrule*[left=T-Or \ ]
  {\typedexp{\Tenv}{b_1}{\mkterm{bool}}{\Tenv'} \\ \typedexp{\Tenv'}{b_2}{\mkterm{bool}}{\Tenv''}}
  {\typedexp{\Tenv}{b_1 \ || \ b_2}{\mkterm{bool}}{\Tenv''}}
  \and
   \inferrule*[left=T-Not \ ]
  {\typedexp{\Tenv}{b}{\mkterm{bool}}{\Tenv'}}
  {\typedexp{\Tenv}{!b}{\mkterm{bool}}{\Tenv''}}
  \end{mathpar}
  
\subsection{Revised typing rules for field and variable dereference}
\begin{mathpar}
    \inferrule*[left=T-LinVar \ ]
  {\linterm(g)}
  {\typedexp{(\Tenv,r:g)}{r}{g}{\Tenv}}
 \and
 \inferrule*[left=T-UnVar \ ]
  {\unterm(g)}
  {\typedexp{(\Tenv,r:t)}{r}{g}{(\Tenv,r:g)}}
 \and
 \inferrule*[left=T-LinField \ ]
  {\typedval{\Tenv}{this}{\objecttype{C}{u;\vec{F}}} \\ 
   \vec{F}(f)=g \\ \linterm(g) 
 }
  {\typedexp{\Tenv}{\der{\mkterm{this}}{f}}{t}{\changetype{\Tenv}{\mkterm{this}}{\objecttype{C}{u;(\vec{F}\setminus f)}}}} 
 \and
 \inferrule*[left=T-UnField \ ]
  {\typedval{\Tenv}{\mkterm{this}}{\objecttype{C}{u;\vec{F}}} \\ 
  \vec{F}(f)=t
    \\ \unterm(t)}
  {\typedexp{\Tenv}{\der{\mkterm{this}}{f}}{t}{\Tenv}} 
  \and
 \inferrule*[left=T-NullField \ ]
  {\typedval{\Tenv}{\mkterm{this}}{\objecttype{C}{u;\vec{F}}} \\ (\_ \ f) \notin \vec{F}}
  {\typedexp{\Tenv}{\der{\mkterm{this}}{f}}{\mkterm{null}}{\Tenv}} 
  \end{mathpar}
  
  \newpage
\subsection{Revised typing rules for simple statements}
\begin{mathpar}
  \inferrule*[left=T-Seq]
  {\typedexp{\Tenv}{s}{t}{\Tenv'} \\ \typedexp{\Tenv'}{t'}{g'}{\Tenv''}}
  {\typedexp{\Tenv}{s;s'}{t'}{\Tenv''}}
  \end{mathpar}
  \begin{mathpar}
	\inferrule*[left=T-AssignVar \ ]
  {e \neq \mkterm{null} \\ \typedexp{\Tenv}{d}{g}{\Tenv'} \\ \typedexp{\Tenv}{e}{g'}{\Tenv'} \\ \agreeterm(g',g)}
  {\typedexp{\Tenv}{d = e}{\unittype}{\Tenv'}}  
	\and
	\inferrule*[left=T-AssignField \ ]
  {e \neq \mkterm{null} \\ \typedexp{\Tenv}{e}{g}{\Tenv'} \\ \typedval{\Tenv'}{\mkterm{this}}{\objecttype{C}{u;\vec{F}}} \\ C.\fieldsterm(f) = g' \\ (\_~f) \notin \vec{F} \vee \vec{F}(f) = g \\ \agreeterm(g',g)}
  {\typedexp{\Tenv}{\upd{\mkterm{this}}{f}{e}}{\unittype}{\changetype{\Tenv'}{\mkterm{this}}{\objecttype{C}{u;(\vec{F}\cup (g~f))}}}}  
	\and
	\inferrule*[left=T-AssignFieldNull \ ]
  {\typedval{\Tenv'}{\typedexp{\Tenv}{e}{\mkterm{null}}{\Tenv} \\ \mkterm{this}}{\objecttype{C}{u;\vec{F}}}}
  {\typedexp{\Tenv}{\upd{\mkterm{this}}{f}{e}}{\unittype}{\changetype{\Tenv'}{\mkterm{this}}{\objecttype{C}{u;(\vec{F} \setminus (\_~f))}}}}  
  \and
  \inferrule*[left=T-NewVar \ ]
  {d \notin \Gamma \\ \typedexp{\Tenv}{e}{g'}{\Tenv'} \\ \agreeterm(g',g)}
  {\typedexp{\Tenv}{g \ d = e}{\unittype}{\changetype{\Tenv'}{d}{g}}} 
  \and
  \inferrule*[left=T-Spawn \ ]
  {\typedexp{\Tenv}{s}{t}{\Tenv'} \\ \unterm(\mkterm{modified}(\Tenv, \Tenv')) \\ \mkterm{completed}(\Tenv, \Tenv')}
  {\typedexp{\Tenv}{\mkterm{spawn} \ \{s\}}{\mkterm{void}}{\Tenv'}}
  \end{mathpar}
  
\subsection{Revised typing rules for control flow expressions}
\begin{mathpar}
   \inferrule*[left=T-IfCall \ ]
   {w \neq \mkterm{this} \\ \typedexp{\Tenv}{w.m(e)}{\booltype}{\Tenv'} \\ \typedval{\Tenv'}{w}{\objecttype{C}{\langle u_t + u_f \rangle ;\vec{F}}} \\ 
   \typedexp{{\changetype{\Tenv'}{w}{\objecttype{C}{u_t;\vec{F}}}}}{s'}{t}{\Tenv''} \\ 
   \typedexp{{\changetype{\Tenv'}{w}{\objecttype{C}{u_f;\vec{F}}}}}{s''}{t}{\Tenv''}}
   {\typedexp{\Tenv}{\cond{w.m(e)}{s'}{s''}}{t}{\Tenv''}} 
   \and
   \inferrule*[left=T-IfNotCall \ ]
   {w \neq \mkterm{this} \\ \typedexp{\Tenv}{w.m(e)}{\booltype}{\Tenv'} \\ \typedval{\Tenv'}{w}{\objecttype{C}{\langle u_t + u_f \rangle ;\vec{F}}} \\ 
   \typedexp{{\changetype{\Tenv'}{w}{\objecttype{C}{u_f;\vec{F}}}}}{s'}{t}{\Tenv''} \\ 
   \typedexp{{\changetype{\Tenv'}{w}{\objecttype{C}{u_t;\vec{F}}}}}{s''}{t}{\Tenv''}}
   {\typedexp{\Tenv}{\cond{!w.m(e)}{s'}{s''}}{t}{\Tenv''}} 
   \and
   \inferrule*[left=T-IfUnCall \ ]
   {\typedexp{\Tenv}{w.m(e)}{\booltype}{\Tenv'} \\ \typedval{\Tenv'}{w}{\objecttype{C}{\epsilon ;\vec{F}}} \\ 
   \typedexp{\Tenv'}{s'}{t}{\Tenv''} \\ \typedexp{\Tenv'}{s''}{t}{\Tenv''}}
   {\typedexp{\Tenv}{\cond{w.m(e)}{s'}{s''}}{t}{\Tenv''}} 
   \and
   \inferrule*[left=T-If \ ]
   {\typedexp{\Tenv}{b}{\booltype}{\Tenv} \\ \typedexp{\Tenv}{s'}{t}{\Tenv''} \\ \typedexp{\Tenv}{s''}{t}{\Tenv''}}
   {\typedexp{\Tenv}{\cond{b}{s'}{s''}}{t}{\Tenv''}} 
   \and
  \and
   \inferrule*[left=T-WhileCall \ ]
   {w \neq \mkterm{this} \\ \typedexp{\Tenv}{w.m(e)}{\booltype}{\Tenv'} \\ \typedval{\Tenv'}{w}{\objecttype{C}{\langle u_t + u_f \rangle ;\vec{F}}} \\ 
    \typedexp{{\changetype{\Tenv'}{w}{\objecttype{C}{u_t;\vec{F}}}}}{s'}{t}{\Tenv''} \\ \Tenv(w) = \Tenv''(w)}
   {\typedexp{\Tenv}{\while{w.m(e)}{s'}}{t}{\changetype{\Tenv''}{w}{\objecttype{C}{u_f;\vec{F}}}}}  
   \and
   \inferrule*[left=T-WhileNotCall \ ]
   {w \neq \mkterm{this} \\ \typedexp{\Tenv}{w.m(e)}{\booltype}{\Tenv'} \\ \typedval{\Tenv'}{w}{\objecttype{C}{\langle u_t + u_f \rangle ;\vec{F}}} \\ 
    \typedexp{{\changetype{\Tenv'}{w}{\objecttype{C}{u_f;\vec{F}}}}}{s'}{t}{\Tenv''} \\ \Tenv(w) = \Tenv''(w)}
   {\typedexp{\Tenv}{\while{!w.m(e)}{s'}}{t}{\changetype{\Tenv''}{w}{\objecttype{C}{u_t;\vec{F}}}}}  
   \and
   \inferrule*[left=T-WhileUnCall \ ]
   {\typedexp{\Tenv}{w.m(e)}{\booltype}{\Tenv'} \\ \typedval{\Tenv'}{w}{\objecttype{C}{\epsilon ;\vec{F}}} \\ 
    \typedexp{\Tenv'}{s}{t}{\Tenv} \\ }
   {\typedexp{\Tenv}{\while{w.m(e)}{s}}{t}{\Tenv'}}  
   \and
  \inferrule*[left=T-While \ ]
   {\typedexp{\Tenv}{b}{\booltype}{\Tenv} \\ \typedexp{\Tenv}{s}{t}{\Tenv}}
   {\typedexp{\Tenv}{\while{b}{s}}{t}{\Tenv}}  
  \end{mathpar}
  
\subsection{Typing rules for subtyping}
\begin{mathpar}
  \inferrule*[left=T-InjL]
  {\typedexp{\Tenv}{e}{t}{\Tenv'}}
  {\typedexp{\Tenv}{e}{t}{\langle \Tenv' + \Tenv'' \rangle}}
  \and
  \inferrule*[left=T-InjR]
  {\typedexp{\Tenv}{e}{t}{\Tenv''}}
  {\typedexp{\Tenv}{e}{t}{\langle \Tenv' + \Tenv'' \rangle}}
  \and
  \inferrule*[left=T-Sub]
  {\typedexp{\Tenv}{e}{C[u]}{\Tenv'} \\ C[u] <: C[u']}
  {\typedexp{\Tenv}{e}{C[u']}{\Tenv'}}
  \inferrule*[left=T-SubEnv]
  {\typedexp{\Tenv}{e}{t}{\Tenv'} \\ \Tenv' <: \Tenv''}
  {\typedexp{\Tenv}{e}{t}{\Tenv'}}
  \end{mathpar}
  
\subsection{Revised typing rules for calls}
\begin{mathpar}
  \inferrule*[left=T-New]
  {\typedexp{\Tenv}{e}{t'}{\Tenv'} \\ \\  C.\usageterm = \mkterm{lin}\{C; u\} \\ (\methodd{t}{C}{t'\, x}{s}, \_) \in C.\methsterm \\ \unterm(\Tenv' \backslash \Tenv)}
  {\typedexp{\Tenv}{\mkterm{new} \ C (e)}{\objecttype{C}{u}}{\Tenv'}}
  \and
  \inferrule*[left=T-UnNew]
  {\typedexp{\Tenv}{e}{t'}{\Tenv'} \\ \\  C.\usageterm = \epsilon \\ (\methodd{t}{C}{t'\, x}{s}, \_) \in C.\methsterm \\ \unterm(\Tenv' \backslash \Tenv)}
  {\typedexp{\Tenv}{\mkterm{new} \ C (e)}{\objecttype{C}{u}}{\Tenv'}}
  \and
      \inferrule*[left=T-SelfCall1 \ ]
  	{\typedexp{\Tenv}{e}{t'}{\Tenv'} \\ \typedval{\Tenv'}{\this}{\objecttype{C}{u;\vec{F}}}\\ (\methodd{t}{m}{t'\, x}{s}, 0) \in C.\methsterm \\ \typedexp{\Tenv'}{s}{t}{\Tenv''} \\ \unterm(\Tenv' \backslash \Tenv)}  
  {\typedexp{\Tenv}{\methcal{\this}{m}{e}}{t}{\Tenv''}}
\and
      \inferrule*[left=T-SelfCall2 \ ]
  	{\typedexp{\Tenv}{e}{t'}{\Tenv'} \\ \typedval{\Tenv'}{\this}{\objecttype{C}{u;\vec{F}}}\\ (\methodd{t}{m}{t'\, x}{s}, 1) \in C.\methsterm}  
  {\typedexp{\Tenv}{\methcal{\this}{m}{e}}{t}{\Tenv'}}
\and
   \inferrule*[left=T-Call \ ]
  {w \neq \mkterm{this} \\ \typedexp{\Tenv}{e}{t'}{\Tenv'} \\ \typedval{\Tenv'}{w}{\objecttype{C}{u;\vec{F}}}\\  u.\allowsterm(m)= u'  \\ (\methodd{t}{m}{t'\, x}{s}, \_) \in C.\methsterm \\ \unterm(\Tenv' \backslash \Tenv)}
  {\typedexp{\Tenv}{\methcal{w}{m}{e}}{t}{\changetype{\Tenv'}{w}{\objecttype{C}{u';\vec{G}}}}}
  \end{mathpar}

\end{document}